\documentclass[%
reprint,
secnumarabic,
amssymb, 
nobibnotes,
aps, 
prd,
superscriptaddress,
nofootinbib,
amsmath,amssymb,aps, showkeys
]{revtex4-1}

\usepackage{graphicx}

\graphicspath{{./figs_v3/}}
\usepackage{dcolumn}
\usepackage{enumitem}
\usepackage{bm}
\usepackage[normalem]{ulem}
\usepackage[mathlines]{lineno}% Enable numbering of text and display math
\usepackage{amsmath}
\usepackage{hyperref}
\hypersetup{
    colorlinks=True,
    linkcolor= blue,
    citecolor= blue,
    filecolor= blue,      
    urlcolor=blue,
}
\usepackage{cleveref}
\usepackage{float}

\begin{document}
\sloppy

\title{Towards a Composite Framework for Simultaneous Exploration of New Physics in Background and Perturbed Universe}

\author{Shibendu Gupta Choudhury}
\email{pdf.schoudhury@jmi.ac.in}
\author{Purba Mukherjee}
\email{pdf.pmukherjee@jmi.ac.in}
\author{Anjan Ananda Sen}
\email{aasen@jmi.ac.in}

\affiliation{Centre for Theoretical Physics, Jamia Millia Islamia, New Delhi-110025, India}


%\pacs{}

%\date{\today}

\begin{abstract}
We investigate deviations from $\Lambda$CDM by independently parameterizing modifications in the background evolution and the growth of structures. The background is characterized by two parameters, $A$ and $B$, which reduce to $A=\Omega_{m0}$ and $B=2/3$ in the $\Lambda$CDM limit, while deviations in the growth of structures are captured through a fitting function for $f\sigma_8$ involving the growth index $\gamma$. { Using recent observational datasets involving background expansion and growth of structures (related to observations involving redshift space distortions), we find significant evidence for departures from $\Lambda$CDM in the background expansion whereas there is no finite evidence for deviations from $\Lambda$CDM behaviour in the growth of structures. This suggests that with the current precision in observational data involving background and perturbed Universe, a deviation from $\Lambda$CDM behaviour is confirmed (as shown in the recent DESI-DR2 results). But whether this deviation is due to an evolving Dark Energy or due to the modification of gravity at cosmological scales is still an open question, largely due to less precise data from perturbed Universe.  We further demonstrate how the future high-precision growth data (from Euclid, for example) can answer such question using a forecast study.}
\end{abstract}

\maketitle
\newpage

\section{Introduction}

The $\Lambda$CDM model, which assumes General Relativity (GR) as the underlying theory of gravity and a spatially flat universe having approximately 70\% Dark Energy (DE) (with constant energy density described by a cosmological constant $\Lambda$) and 30\% Cold Dark Matter (CDM), provides an excellent agreement with observations \cite{Peebles:2024txt}. This includes fluctuations in the temperature and polarization of the cosmic microwave background (CMB) \cite{Planck:2018vyg, Planck:2015bue, ACT:2020gnv, ACT:2025fju}, observations involving the large-scale structure (LSS) of the universe \cite{BOSS:2014hhw, BOSS:2016wmc,Addison:2017fdm,Haridasu:2017ccz},  the distance-redshift relation of Type Ia supernovae (SNIa) \cite{Riess:2019qba,Pan-STARRS1:2017jku}, etc. However, recent observations suggest a significant deviation from this concordance model. For example, when baryon acoustic oscillation (BAO) measurements from the Dark Energy Spectroscopic Instrument (DESI) \cite{DESI:2024mwx, DESI:2025zgx} are combined with SNIa \cite{Rubin:2023ovl,DES:2024jxu,Brout:2022vxf} and CMB \cite{Planck:2018vyg, ACT:2020gnv} observations, it indicates $\approx 3-4\sigma$ deviation from $\Lambda$CDM model).
Furthermore, several theoretical challenges persist, such as the elusive nature of dark matter, the cosmological constant problem, and the coincidence problem \cite{Weinberg:2000yb}. These unresolved issues provide strong motivation to explore potential deviations from $\Lambda$CDM and investigate alternative cosmological frameworks \cite{DiValentino:2021izs, Abdalla:2022yfr}.

Several approaches have been proposed so far to model deviations from $\Lambda$CDM. They can be broadly classified into two scenarios: (i) modification in the DE sector or (ii) modification in the gravity sector. In the first class, $\Lambda$ is replaced by a component for which the equation of state (EoS) for DE $w_\mathrm{DE}$ is either constant (but $\neq -1$) or has a scale-factor dependence \cite{Chevallier:2000qy,Linder:2002et}. The latter represents a large family of dynamical dark energy models (DDE) \cite{Linder:2002et,Linder:2005in,Capozziello:2005ra} and GR is assumed to be the governing theory in these models. The other class represents models where late-time acceleration of the Universe occurs due to modification of gravity at cosmological scales. This includes modified gravity (MG) theories like $f(R)$ gravity \cite{Sotiriou:2008rp,Nojiri:2017ncd}, scalar-tensor theories \cite{Yasunori_Fujii_2003}, Cardassian models \cite{Freese:2002sq} and other alternatives \cite{Dvali:2000hr, Barreiro:2004bd}.

In recent years, reconstruction methods have become increasingly popular in cosmology \cite{Sahni:2006pa}. These include non-parametric techniques to infer the underlying cosmic evolution directly from data \cite{Holsclaw:2010sk, Holsclaw:2011wi, Sahlen:2005, Huterer:2003} or parametrization(s) that can introduce flexible yet physically motivated modifications to key cosmological functions \cite{Starobinsky:1998fr, Huterer:1998qv, Chevallier:2000qy, Sen:2001xu, Sen:2005ra}. In this work, we adopt the latter for examining deviations from $\Lambda$CDM in both the background evolution and the growth of cosmic structures. Rather than adopting a single unified modification (as studied in \cite{Huterer:2022dds, Nguyen:2023fip}), we attempt to model deviations separately in the background and perturbation sectors. This framework allows us to explore potential new physics in each sector independently, {\color{black} and to study both DDE and MG scenarios in a unified way without focusing on specific models (e.g., particular cases of $f(R)$-gravity or DDE, as discussed in \cite{Odintsov:2024woi}).} To capture deviations in the background sector, we adopt the formalism discussed in \cite{Sen:2001xu, Mukhopadhyay:2024fch}, where modifications are introduced in the late-time evolution of the scale factor $a(t)$, the most fundamental quantity governing cosmological evolution. { This is primarily motivated from the fact that almost all the observables related to background expansion are directly related to the total energy density of the Universe ($\rho_{t}$) which in-turn is related to the time derivative of the scale factor $a(t)$ through Einstein equations. Hence the observables in the background expansions are completely agnostic to whether the $\rho_{t}$ represents a dark energy which can be minimal or non-minimal or whether $\rho_{t}$ represents a modified gravity model. As shown in \cite{Mukhopadhyay:2024fch}, the general formalism is quite robust to account the deviations in the background evolution from $\Lambda$CDM behaviour for both DE and MG scenarios. 

Next, we also need to model the growth of structures (at the level of linear fluctuations) in a model agnostic way. This has been already studied in the literature  \cite{Huterer:2022dds, Nguyen:2023fip} where the growth factor $f$ for the linear matter fluctuations has been modeled as $f\sim \Omega_{m}^{\gamma}(z)$ \cite{Fry:1985zy, Wang:1998gt, Linder:2005in} where $\Omega_{m}(z)$ represents the evolution of the matter density parameter. For $\Lambda$CDM model, $\gamma = 6/11$ and hence any deviation from $\gamma =6/11$ with observational data (for e.g. redshift space distortions) can signal new physics in the perturbed sector. As the growth data from redshift space distortion involves the combination $f\sigma_{8}$ ($\sigma_8$ being the variance of matter density fluctuations in a sphere of comoving radius 8 $h^{-1}$ Mpc) instead, we first construct a fitting function for $f\sigma_{8}$ that can identify the deviation from $\Lambda$CDM model in the perturbed Universe using the data from redshift space distortion.}
 
 { With these two model independent prescriptions for background evolution and growth of matter fluctuations, subsequently we constrain them} by considering low redshift SNIa and BAO data, along with growth rate measurements from redshift space distortions (RSD).

{\color{black} This paper is organized as follows: In Section \ref{theory}, we present the underlying theoretical framework. Section \ref{data} briefly outlines the datasets and methodology used in our analysis, while Section \ref{results} discusses the results in detail. In Section \ref{spmd}, we show that our general framework can accommodate several specific models, including both DDE and MG. Section \ref{fc} explores how upcoming, more precise RSD data can further improve our understanding of cosmic evolution. Finally, Section \ref{conc} contains our concluding remarks.}

\section{Theoretical Framework}\label{theory}

We know that for the $\Lambda$CDM model, the late time behaviour of the scale factor of the Universe is given by,
\begin{equation}
 a(t)=a_1 \, [\sinh (t/\tau)]^{\frac{2}{3}} \, .
\end{equation}
Here $a_1$ is an arbitrary dimensionless parameter, and $\tau$ is an arbitrary parameter of dimension [{\it T }].
The resulting Hubble parameter $H(z)$ is,
\begin{equation}
 \label{eqn:H_LCDM}
H(z) = H_{0} \, [\Omega_{m0} (1+z)^{3} + (1-\Omega_{m0})]^{\frac{1}{2}} \, ,
\end{equation}
where $H_0$ is the current value of the Hubble parameter and $\Omega_{m0}$ is the current fractional matter density parameter given by, {\color{black} $\Omega_{m0}=\frac{\rho_{m0}}{\rho_{c0}}$, where $\rho_{c0}=3H_{0}^2/8\pi G$ is the critical density of the Universe at current times.}
We define the reduced Hubble factor as,
\begin{equation}\label{lcdmE}
 E(z)\equiv {H(z)}/{H_0}= \sqrt{\Omega_{m0} (1+z)^{3} + (1-\Omega_{m0})} \, .
\end{equation}
The fractional matter density parameter $\Omega_{m}$  at some redshift $z$ is given by,
\begin{equation}\label{Ommz}
 \Omega_{m}(z)=\frac{\Omega_{m0}(1+z)^3}{E^2(z)}.
\end{equation}

Next, we consider a minimal extension of $a(t)$ to represent any new physics dominating the background evolution, irrespective of any specific DE model or MG theory \cite{Sen:2001xu, Mukhopadhyay:2024fch},
\begin{equation}\label{eqn:sf_ABCDM}
a(t) = \tilde{a}_{1} [\sinh (t/\tau)]^{B} \, , 
\end{equation}
where $\tilde{a}_{1} $ and $B$ are dimensionless arbitrary parameters.  Accordingly, the expression for $E(z)$ comes out to be,
\begin{equation}\label{eqn:H_ABCDM}
E(z) = \left[A (1+z)^{2/B} + (1-A)\right]^{\frac{1}{2}} \, .
\end{equation}
Here, $H_{0} = \frac{B}{\tau \sqrt{1-A}}$ is the Hubble parameter at present epoch, and  $A = \frac{\tilde{a}_{1}^{2/B}}{1+\tilde{a}_{1}^{2/B}}$. We can retrieve the $H(z)$ for $\Lambda$CDM with the identification $A=\Omega_{m0}$ and $B=2/3$ in Eq. \eqref{eqn:H_ABCDM}. This modified evolution can have several origins that are explicitly discussed in \cite{Mukhopadhyay:2024fch} as well as situations where there is a running in the cosmological parameters \cite{Colgain:2024mtg,Malekjani:2023ple}. However, in this work, we do not focus on these origins. Instead, we assume that the parameters $A$ and $B$ characterize a generic framework for modified background evolution and encode any potential new physics in the background sector.\\

Similarly, any new physics in the perturbation sector can be treated distinctly, as we describe next. At late times, the evolution of matter density contrast $\delta = \delta \rho_m / {\rho_m}$ within the framework of $\Lambda$CDM is governed by the following equation,
\begin{equation}\label{greqlcdm}
 \delta^{\prime \prime}+\frac{1}{a}\left(3+\frac{a \, E^\prime}{E}\right)\delta^\prime-\frac{3 \, \Omega_{m0}}{2 \, a^5 \, E^2}\delta=0 \, ,
\end{equation}
where $\prime$ denotes differentiation with respect to $a$. Note that the presence of $\Omega_{m0}$ (the matter contribution in the Universe) in the above equation originates from Poisson's equation, where we assume that only matter clusters at relevant scales. This is important because even if the DE sector contains a term which scales as $\sim (1+z)^3$, that will not contribute to Poisson's equation. On the other hand, the term $\frac{E^{\prime}}{E}$ in the above equation contains contributions from both matter and DE through Eq. \eqref{eqn:H_ABCDM}.

A robust parametrization for the logarithmic growth factor $f\equiv \frac{\mathrm{d}\ln \delta}{\mathrm{d}\ln a}$ is given by \cite{Fry:1985zy, Wang:1998gt, Linder:2005in}, 
\begin{equation}\label{ffit}
f=\Omega_{m}^\gamma \, ,                                          
\end{equation}
where $\gamma$ is the growth index parameter. With GR as the underlying theory, flat $\Lambda$CDM background predicts $\gamma=6/11$, and this fit is accurate to 0.1\% \cite{Linder:2005in, Linder:2007hg, Gong:2008fh}. Note that in the concordance  $\Lambda$CDM  model, the linear matter density fluctuation $\delta$ is governed solely by the background evolution. Therefore, any deviation from $\gamma = 6/11$ would indicate a potential new physics affecting the growth of structure beyond what is dictated by background evolution alone. 

As our first step, we construct a suitable fitting form for $f\sigma_8$ as a function of $\Omega_m$ to introduce new physics in the perturbation sector. The primary motivation comes from the fact that observational data usually involve $f\sigma_8$ rather than $f$ alone. A well-constructed fitting form will thus offer a more convenient way to introduce modifications in the perturbation sector while keeping the background sector unaffected.

\begin{figure}[t]
\includegraphics[width=\linewidth]{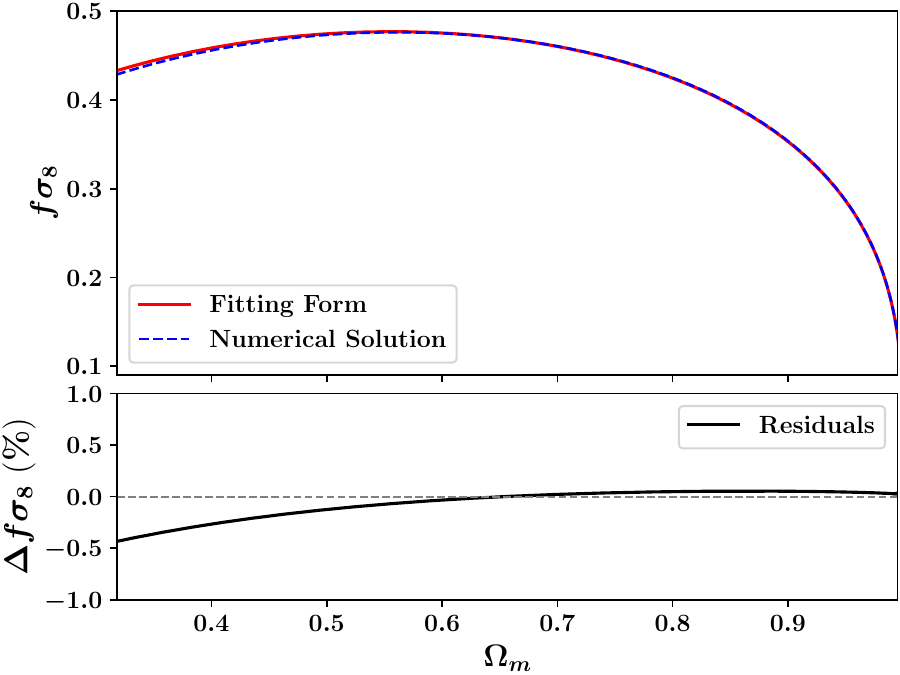}
 \caption{Plots for $f\sigma_8$ values as a function of $\Omega_m$ obtained using the fitting form in Eq. \eqref{fs8fit} (red line) and from solving Eq. \eqref{greqlcdm} (dashed blue line). Here we have used $\Omega_{m0}=0.315,~ \sigma_{8,0}=0.811, \gamma=\frac{6}{11}$ and assumed $\delta(a)\approx a$ and $\delta'(a)\approx 1$ in the matter dominated era.}\label{fig1}
\end{figure}

The evolution of $\sigma_8$ is defined as $\sigma_8 = \sigma_{8,0} \, \frac{\delta(a)}{\delta(1)}$, where $\sigma_{8,0}$ is its present value. To express $\sigma_8$ as a function of $\Omega_{m}$ we note,
\begin{equation}
 \frac{d \log \delta}{d \Omega_m}=\frac{\Omega_m^{\gamma-1}}{3(\Omega_m-1)} \, ,
\end{equation}
as derived from Eqs. \eqref{lcdmE}, \eqref{Ommz}, and \eqref{ffit}. Thus, we arrive at the solution for $\delta$, as
\begin{equation}
 \delta(\Omega_m)=A \, \exp\left[-\frac{\Omega_m^{\gamma } \, _2F_1(1,\gamma ;\gamma +1;\Omega_m)}{3 \gamma }\right] \, ,
\end{equation}
where $\, _2F_1({\rm a}, \, {\rm b}; \, {\rm c}; \, {\rm x})$ is the Gauss Hypergeometric function. Therefore, 
\begin{equation}\label{fs8fit}
\begin{split}
f\sigma_8(\Omega_m)=\sigma_{8,0} \, \Omega_m^\gamma  &  \exp \left[\frac{\Omega_{m0}^{\gamma }}{3\gamma} \, _2F_1(1,\gamma ;\gamma +1;\Omega_{m0})\right.  \\  &  \left. -  \frac{\Omega_m^{\gamma }}{3\gamma} \, _2F_1(1,\gamma ;\gamma +1;\Omega_m)\right].
 \end{split}
\end{equation}
 The above expression provides a new fitting form for $f\sigma_8$ as a function of $\Omega_m$. Note that equations (9)-(11) are only valid for $\Lambda$CDM model with $\gamma = 6/11$. This function for $f\sigma_{8}$(given by equation (11)) matches very well with the $f\sigma_8$ values obtained by solving Eq. \eqref{greqlcdm} for the $\Lambda$CDM model. Fig. \ref{fig1} illustrates this agreement by comparing the fitting function in \eqref{fs8fit} with the numerical solution of the growth equation (Eq. \eqref{greqlcdm}). As evident from Fig. \ref{fig1}, this fitting function describes the $f\sigma_{8}$ behaviour for $
 \Lambda$CDM model very accurately. With this, any new physics in the perturbation sector will be captured by the parameter $\gamma\neq \frac{6}{11}$.

In what follows, we utilize Eq. \eqref{eqn:H_ABCDM} to represent the background evolution and Eq. \eqref{fs8fit} with a general $\gamma\neq \frac{6}{11}$ to represent the growth of perturbations. This defines our composite model with parameters $A$, $B$, $\Omega_{m0}$, $\gamma$ and $\sigma_{8,0}$, which we constrain employing different observational datasets and their combinations. In the $\Lambda$CDM model, $B=2/3$, $\gamma = 6/11$ and $A \equiv \Omega_{m0}$. Deviations from any of these conditions would indicate potential new physics in either background or in the perturbed Universe or in both.

\begin{table*}[t]
\begin{center}
\renewcommand{\arraystretch}{1.35}
\setlength{\tabcolsep}{10pt}
\resizebox{\textwidth}{!}{
\begin{tabular}{l c c c c c c c c}
\hline\hline
\textbf{Datasets} & $\boldsymbol{hr_d}$ & $\boldsymbol{A}$ & $\boldsymbol{B}$ & $\boldsymbol{\Omega_{m0}}$ & $\boldsymbol{\gamma}$ & $\boldsymbol{\sigma_{8,0}}$ & $\boldsymbol{H_0}$ & $\boldsymbol{S_8}$ \\
\hline
RSD & - & $0.43^{+0.20}_{-0.23}$ & $0.605^{+0.080}_{-0.10}$ & $0.33\pm 0.16$ & $0.95^{+0.34}_{-0.60}$ & $1.01^{+0.13}_{-0.26}$ & - & $1.02\pm 0.32$ \\

DESI-DR2 & $100.6\pm 1.3 $ & $0.331^{+0.034}_{-0.041}$ & $0.684\pm 0.019$ & - & - & - & $68.38\pm 0.87$ & - \\

DESI-DR2+PP & $100.00\pm 0.84$ & $0.348\pm 0.025$ & $0.693\pm 0.014$ & - & - & - & $67.99\pm 0.57$ & - \\

DESI-DR2+DES-5YR & $99.23\pm 0.80$ & $0.373\pm 0.026$ & $0.704\pm 0.014$ & - & - & - & $67.46\pm 0.55$ & - \\

DESI-DR2+RSD & $100.6\pm 1.3$ & $0.328^{+0.033}_{-0.041}$ & $0.683\pm 0.019$ & $0.220^{+0.11}_{-0.055}$ & $0.54\pm 0.15$ & $0.871^{+0.052}_{-0.11}$ & $68.42\pm 0.87$ & $0.712^{+0.18}_{-0.092}$ \\

PP+RSD & - & $0.374^{+0.062}_{-0.099}$ & $0.714^{+0.072}_{-0.11}$ & $0.219^{+0.11}_{-0.076}$ & $0.54^{+0.14}_{-0.18}$ & $0.878^{+0.054}_{-0.12}$ & - & $0.72^{+0.17}_{-0.12}$ \\

DES-5YR+RSD & - & $0.421^{+0.084}_{-0.12}$ & $0.738^{+0.092}_{-0.13}$ & $0.228^{+0.11}_{-0.075}$ & $0.55^{+0.15}_{-0.20}$ & $0.874^{+0.056}_{-0.12}$ & - & $0.73^{+0.17}_{-0.12}$ \\

DESI-DR2+PP+RSD & $100.04\pm 0.85$ & $0.346\pm 0.025$ & $0.691\pm 0.014$ & $0.220^{+0.11}_{-0.059}$ & $0.54\pm 0.15$ & $0.877^{+0.053}_{-0.12}$ & $68.01\pm 0.58$ & $0.712^{+0.18}_{-0.091}$ \\

DESI-DR2+DES-5YR+RSD & $99.23\pm 0.80$ & $0.372\pm 0.025$ & $0.703\pm 0.014$ & $0.228^{+0.11}_{-0.055}$ & $0.55\pm 0.15$ & $0.874^{+0.053}_{-0.11}$ & $67.46\pm 0.55$ & $0.728^{+0.17}_{-0.089}$ \\
\hline\hline
\end{tabular}
}
\end{center}
\caption{Marginalized $1\sigma$ C.L. limits of the parameters of the composite model across various dataset combinations. $H_0$ (in km Mpc$^{-1}$ s$^{-1}$) is derived from the inverse-distance-ladder method assuming $r_d = 147.09 \pm 0.26$ Mpc \cite{Planck:2018vyg}. $S_8$ is calculated as $S_8 = \sigma_{8,0} \sqrt{\Omega_{m0} / 0.3}$.}
\label{Table_gm}
\end{table*}

\begin{figure}[t]
    \centering
    \includegraphics[width=\linewidth]{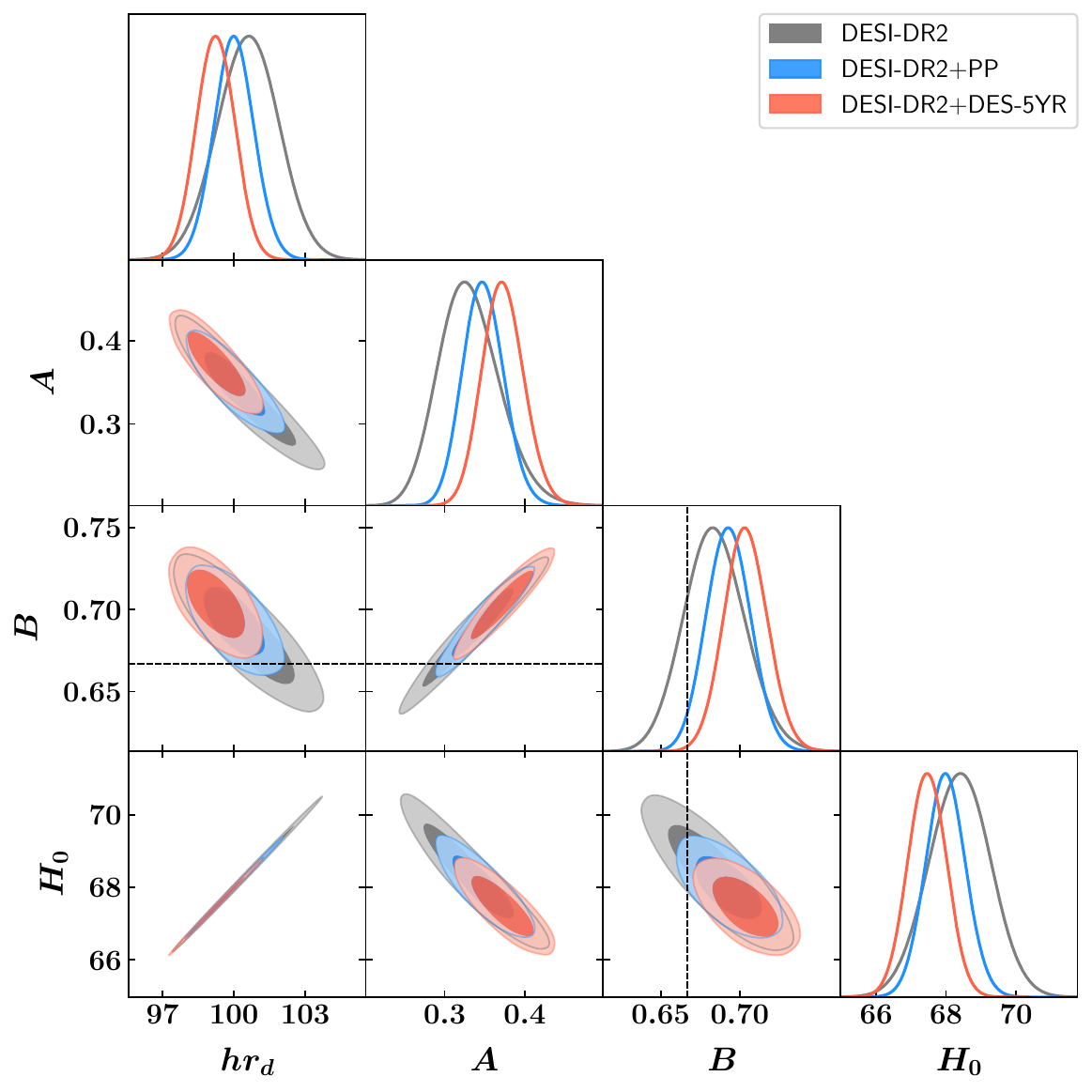}
    \caption{One-dimensional marginalized posterior distributions, 2-dimensional contour plots at 68\% and 95\% C.L. limits for relevant parameters of the composite model using BAO and SNIa datasets.}
    \label{figcombback}
\end{figure}

\begin{figure}[t]
    \centering
     \includegraphics[width=\linewidth]{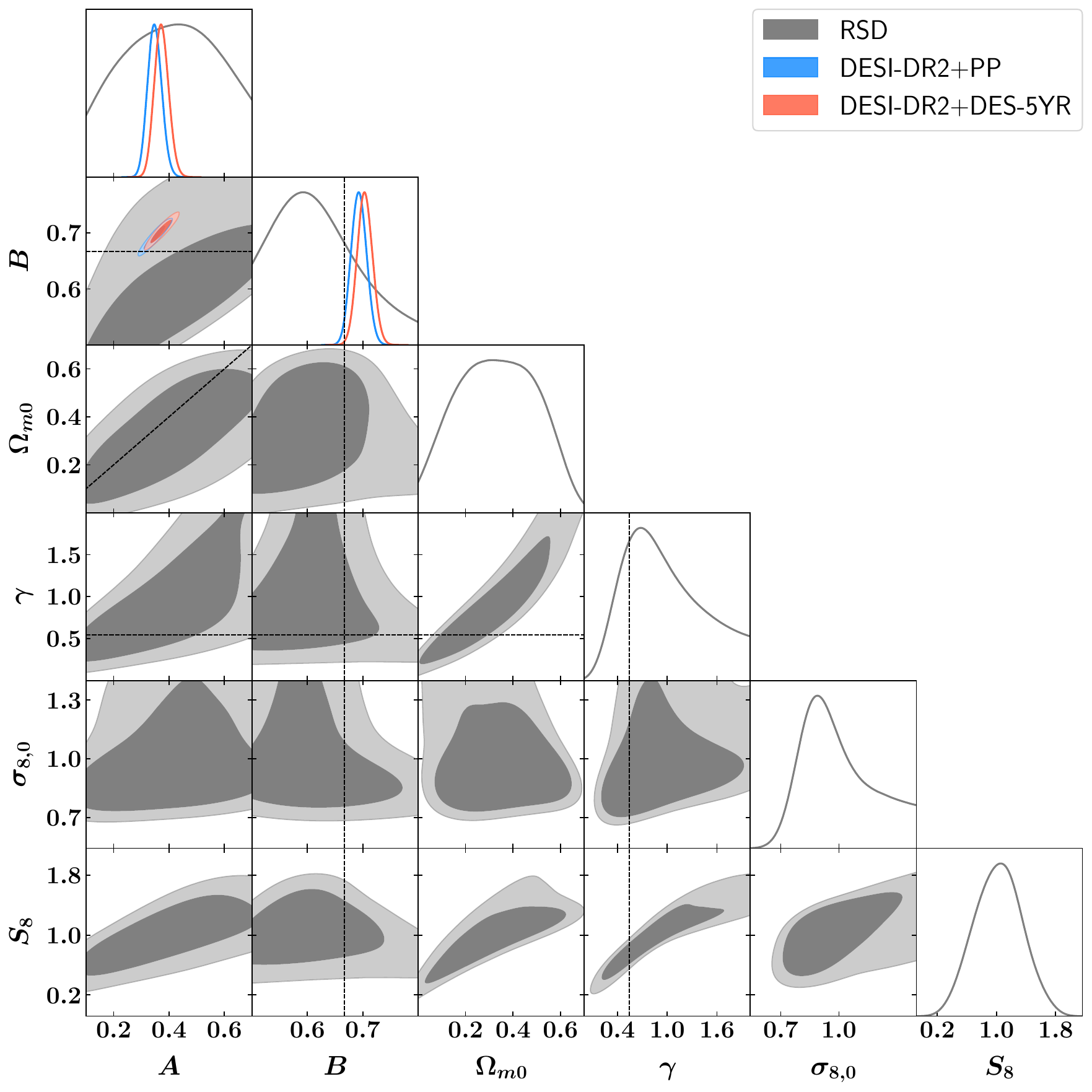}
    \caption{One-dimensional marginalized posterior distributions, 2-dimensional contour plots at 68\% and 95\% C.L. limits for parameters of the composite model using RSD data.}
    \label{figfs8}
\end{figure}

\begin{figure}[t]
    \centering
    \includegraphics[width=\linewidth]{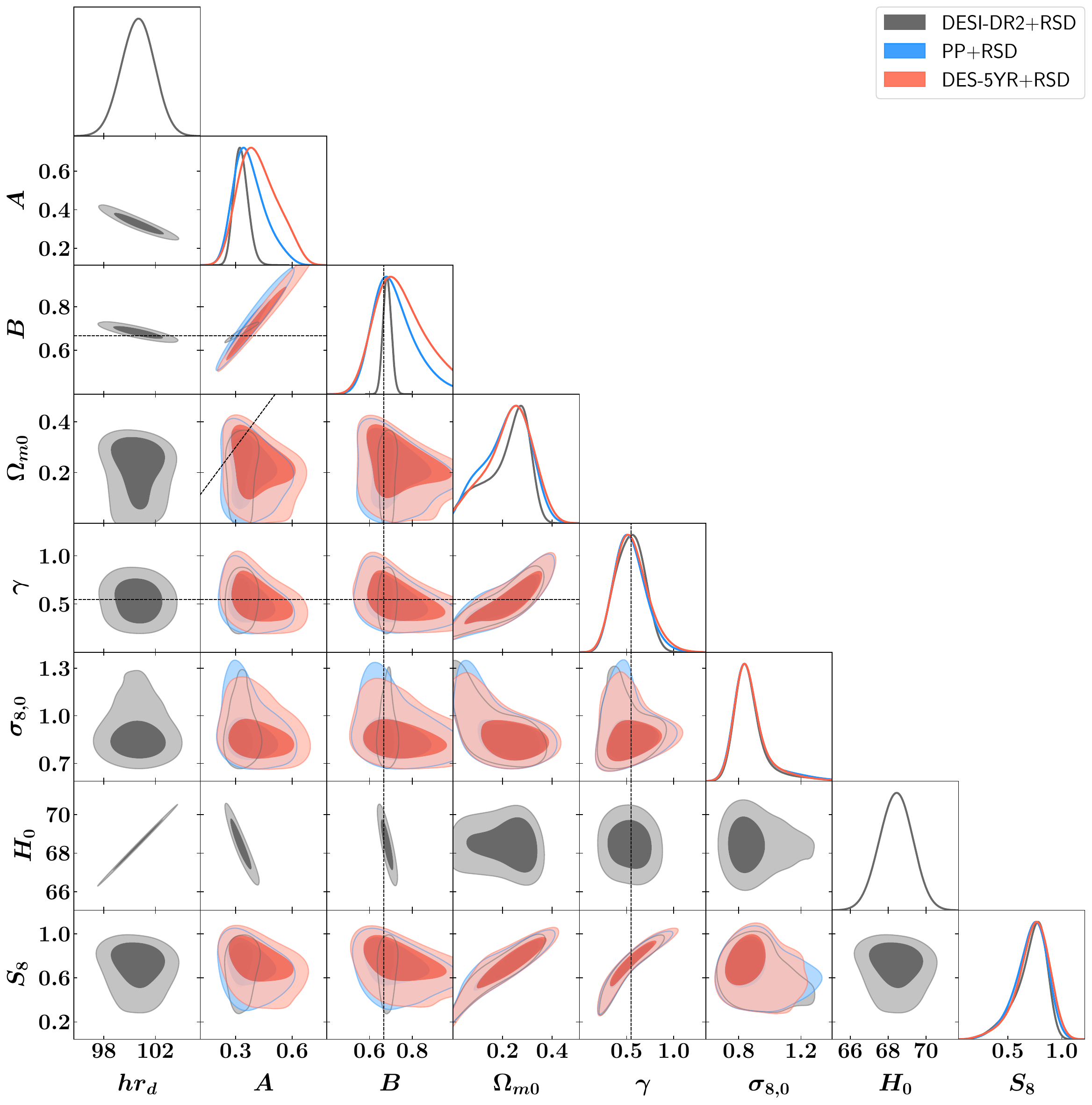}
    \caption{One-dimensional marginalized posterior distributions, 2-dimensional contour plots at 68\% and 95\% C.L. limits for parameters of the composite model using RSD in combination with BAO/SNIa data.}
    \label{figcomb}
\end{figure}

\begin{figure}[t]
    \centering
    \includegraphics[width=\linewidth]{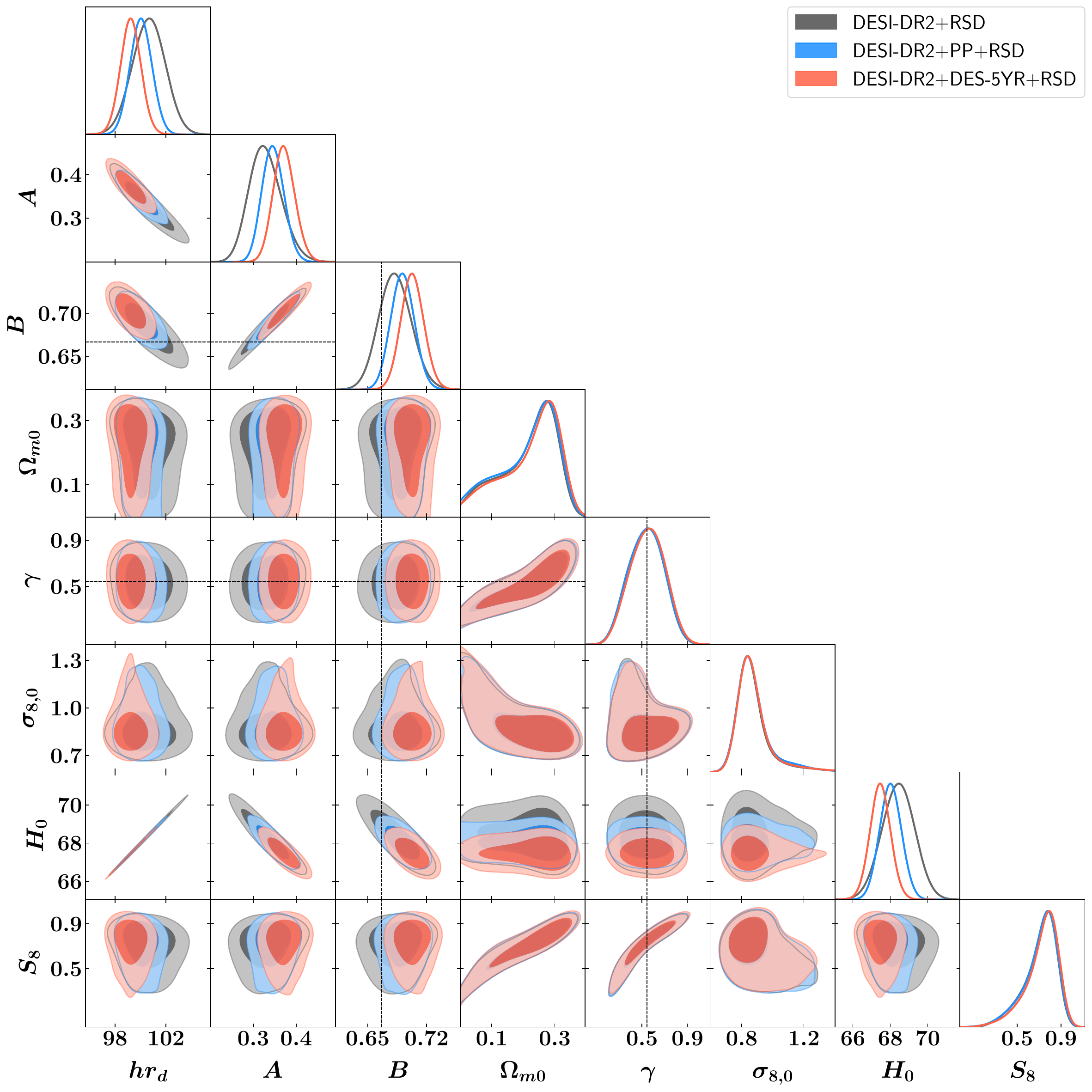}
     \caption{One-dimensional marginalized posterior distributions, 2-dimensional contour plots at 68\% and 95\% C.L. limits for parameters of the composite model using combined BAO, SNIa and RSD datasets.}
     \label{figallcomb}
\end{figure}

\section{Data and Methodology}\label{data}
\noindent We make use of the following datasets in our analysis:
\begin{itemize}[left=0pt]
    \item {\bf BAO}: { We utilize 13 correlated BAO distance measurements from DESI public data release 2 spanning the redshift range $0.1 < z < 4.2$,  assembled in Table IV of \cite{DESI:2025zgx}. For convenience, we refer to this dataset as `DESI-DR2' hereafter.}

    \item {\bf SNIa}: We consider two SNIa datasets, namely the Pantheon-Plus \cite{Brout:2022vxf} compilation and sample provided by 5 years of the Dark Energy Survey (DES) Supernova program \cite{DES:2024upw}. We refer to them as `PP' and `DES-5YR', respectively. 

    \item {\bf RSD}: We consider the $f\sigma_8$ data compiled by Nesseris {\it et al.} \cite{Nesseris:2017vor}, Sagredo {\it et al.} \cite{Sagredo:2018ahx} and Skara and Perivolaropoulos \cite{Skara:2019usd} and include the covariance matrices of the data from the WiggleZ \cite{Blake:2012pj} and SDSS-IV \cite{eBOSS:2018yfg} compilation. We also incorporate the correction from the Alcock-Paczynski (AP) effect by considering the corresponding fiducial cosmology \cite{Macaulay:2013swa}. We refer to this dataset as `RSD'. 
\end{itemize}

Here we have two parameters in the background sector, namely, $A$ and $B$, along with $h r_d$, where $h=\frac{H_0}{100 \text{ km s}^{-1} \text{ Mpc}^{-1}}$ is the dimensionless Hubble constant and $r_d$ is the sound horizon at the drag epoch. In the perturbation sector, we have three more parameters: $\Omega_{m0},~\gamma$ and $\sigma_{8,0}$, in addition to $A$ and $B$. We undertake a Bayesian Markov Chain Monte Carlo (MCMC) analysis to constrain the composite model, utilizing the \texttt{emcee}\footnote{Available at: \href{http://dfm.io/emcee/current/}{http://dfm.io/emcee/current/}} package. { We assume uniform priors on all parameters- $A \in \mathcal{U}{(0,0.9]}$, $B\in \mathcal{U}[0.4,0.85]$, $hr_d\in \mathcal{U}{(80,150)}$, $\Omega_{m0}\in \mathcal{U}(0, 0.7]$, $\gamma\in\mathcal{U}(0,2]$ and $\sigma_{8,0}\in\mathcal{U}[0.4, 1.5]$}. The post-processing of MCMC chains is done using \texttt{GetDist}\footnote{Availalbe at: \href{https://getdist.readthedocs.io/}{https://getdist.readthedocs.io/}}.

\section{Results}\label{results}

\begin{figure}[t]
    \centering
    \includegraphics[width=0.85\linewidth]{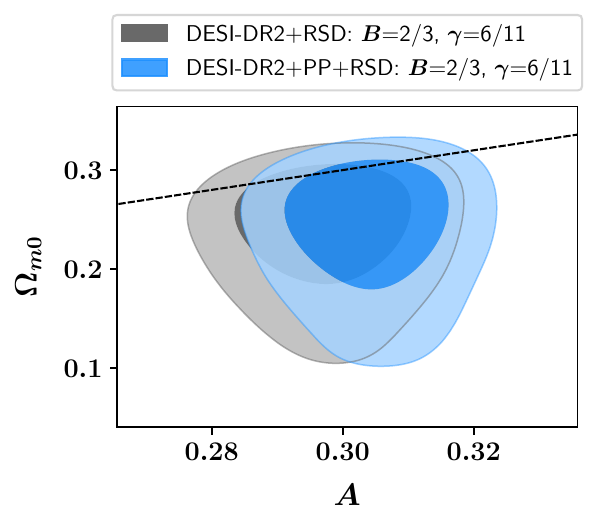}
     \caption{2-dimensional contour plots for $A$ vs $\Omega_{m0}$ at 68\% and 95\% C.L. limits imposing parameters $B=2/3$ and $\gamma = 6/11$ for the composite model using combined BAO, SNIa and RSD datasets.}
     \label{fig:AvsOm}
\end{figure}

The results of our analysis are summarized in Table \ref{Table_gm}, which presents the marginalized constraints on the parameters of the composite model, obtained from different dataset combinations. This table provides an overview of how various observational probes contribute to constraining the parameters governing the background evolution $\left\lbrace hr_d, \, A, \, B\right\rbrace$ and growth rate of cosmic structures $\left\lbrace A, \, B, \, \Omega_{m0}, \, \gamma, \, \sigma_{8,0}\right\rbrace$ respectively. Our findings can be summarized as follows:
\begin{enumerate}[left=0pt]
\item When considering only background evolution datasets such as { DESI-DR2+PP and DESI+DES-5YR}, Fig. \ref{figcombback} shows that the parameters $A$ and $B$ are well constrained at 1$\sigma$ confidence level (CL). Specifically, for the { DESI-DR2+PP dataset, we obtain $A = 0.348\pm{0.025}$ and $B = 0.693 \pm 0.014$, while for the DESI-DR2+DES-5YR dataset, we find $A = 0.373\pm{0.026}$ and $B = 0.704 \pm 0.014$. Note that, for $\Lambda$CDM background evolution, $B = 2/3$. Thus, our obtained values indicate deviations from $\Lambda$CDM at $1.9\sigma$ for DESI-DR2+PP, at $2.7\sigma$ for DESI-DR2+DES-5YR respectively}.

\item Upon incorporating the RSD data to analyze growth constraints, we find that RSD alone does not impose stringent constraints on the relevant parameters. Fig. \ref{figfs8} shows that both $B = 2/3$ and $\gamma = 6/11$ are well within the 1$\sigma$  CL. Additionally, we plot the $A = \Omega_{m0}$ line which illustrates that, with RSD data alone, the equivalence between $A$ and $\Omega_{m0}$ holds within 1$\sigma$ CL. However, the marginalized posteriors for $A$, $\Omega_{m0}$, $\gamma$, and $\sigma_{8,0}$ exhibit non-Gaussian distributions, possibly indicating asymmetries, skewness, or multimodal behaviour in the parameter space. { This is due to the fact that} current RSD measurements are significantly less precise compared to existing BAO and SNIa observations.

\item The constraints on the parameters improve when the RSD data is combined with BAO and/or SNIa datasets, as depicted in Fig. \ref{figcomb}.  { The DESI-DR2+RSD combination yields $A=0.328^{+0.033}_{-0.041}$, $B=0.683\pm 0.019$, $\gamma=0.54\pm 0.15$, $\Omega_{m0}=0.22^{+0.11}_{-0.055}$, and $\sigma_{8,0}=0.871^{+0.052}_{-0.11}$. While the parameters $A$, $B$ and $\gamma$ remain consistent with their $\Lambda$CDM counterparts, best-fit value of $\Omega_{m0}$ shows a noticeable deviation from the Planck prediction. Nevertheless, this deviation remains within the $1\sigma$ CL.}  We also plot the $A=\Omega_{m0}$ line, which reveals that a small but finite region in the $A-\Omega_{m0}$ parameter space accommodates their equivalence within 1$\sigma$ CL. 

\item { Combining the SNIa datasets (PP or DES-5YR) with RSD also results in shifts in the best-fit parameter values relative to their $\Lambda$CDM counterparts. However, for both the PP+RSD and DES-5YR+RSD combinations, the parameters $A$, $B$, and $\Omega_{m0}$ remain consistent with their respective $\Lambda$CDM values within the $1\sigma$ CL, while the best-fit value of $\gamma$ shows excellent agreement with $\Lambda$CDM expectations. }These findings highlight that BAO and SNIa datasets, when used alongside RSD, significantly improve constraints on the perturbation sector.

\item Finally, when combining all three observational probes—BAO, SNIa, and RSD—the constraints become even more refined. The outcomes, presented in Fig. \ref{figallcomb}, indicate that with { DESI-DR2+PP+RSD and DESI-DR2+DES-5YR+RSD combinations, the standard $\Lambda$CDM value of $B = 2/3$ falls outside the $1.8\sigma$ and $2.6\sigma$ CL, respectively}. Nonetheless, the best-fit values of $\gamma$, in both cases, remain in excellent agreement with the theoretical expectation of $\gamma = 6/11$.

\item Given that both $B=2/3$ and $\gamma = 6/11$ (which correspond to $\Lambda$CDM model) are consistent with observational data within $2\sigma$ { for DESI-DR2+RSD and DESI-DR2+PP+RSD}, we explicitly check whether $\Omega_{m0} = A$ holds under these assumptions { for these dataset combinations}. We show in Fig. \ref{fig:AvsOm} the constrained region in $\Omega_{m0}-A$ parameter space assuming $B=2/3$ and $\gamma=6/11$. { As evident from the figure, the $\Omega_{m0} = A$ line lies within a small region of the $1\sigma$ contours. This implies only mild evidence for a non-equivalence between the matter density inferred from the geometric sector and that from the perturbation sector when $B = 2/3$ and $\gamma = 6/11$}. This result is consistent with the findings in \cite{Andrade_2021} where the equivalence between these two matter densities is shown except for some specific dark energy behaviours.

\item The low-redshift observational data that we use in this analysis do not directly constrain the $H_{0}$ parameter but the combination $h r_{d}$ through the BAO measurements. We use the Planck-2018 prior on $r_{d}$ assuming that there is no new physics in the early Universe and obtain the constraint on $H_{0}$ from different dataset combinations. { The resulting constraints on $H_0$ are presented in Table~\ref{Table_gm}.  For the DESI-DR2+RSD combination, we obtain $H_0 = 68.42 \pm 0.87$ km Mpc$^{-1}$ s$^{-1}$}, which naturally accommodates $H_0$ from Planck ($67.4 \pm 0.5$ km Mpc$^{-1}$ s$^{-1}$\cite{Planck:2018vyg}), { and shows a $\sim3.4\sigma$ tension with the SH0ES measurement ($73.04 \pm 1.04$ km Mpc$^{-1}$ s$^{-1}$\cite{Riess:2021jrx}). When SNIa datasets are included, the constraints become more precise as shown in Fig \ref{figallcomb}, thereby increasing the statistical significance of the tension with SH0ES.}

\item For the $\sigma_{8,0}$ parameter, we observe that the marginalized posteriors are non-Gaussian with extended tails to the right. { The mean values of $\sigma_{8,0}$ are consistent with the Planck 2018 \cite{Planck:2018vyg}, ACT-DR6 \cite{ACT:2025fju}, DES-Y3 \cite{DES:2025xii} and KiDS-Legacy \cite{Wright:2025xka} best-fit values within $1\sigma$}. On the other hand, the $S_8$ parameter shows constrained mean values that are { lower to those obtained from the CMB or weak lensing measurements}, mainly due to a reduced mean value of $\Omega_{m0}$. This reduction arises from separating the background and perturbation sectors, introducing a non-equivalence between $A$ (in the background universe) and $\Omega_{m0}$ (in the growth of structures). { Nonetheless, our constraints on $S_8$ are well within 1$\sigma$ of the Planck 2018 \cite{Planck:2018vyg}, ACT-DR6 \cite{ACT:2025fju} CMB and DES-Y3 \cite{DES:2025xii}, KiDS-Legacy \cite{Wright:2025xka} weak lensing results}. 

\item The existing RSD data alone do not provide precise constraints on all parameters (see Fig. \ref{figfs8}) when compared to the background datasets from { DESI-DR2+PP and DESI-DR2+DES-5YR}. However, the inclusion of { DESI-DR2} leads to more precise constraints on parameters like $h r_d$, $A$, and $B$, with these constraints becoming even tighter when { DESI-DR2} is combined with PP and DES-5YR data (see Fig. \ref{figallcomb}). Once $A$ and $B$ are well-constrained, the RSD data then refines the remaining parameters, $ \Omega_{m0} $, $ \sigma_{8,0} $, and $ \gamma $, within the parameter space defined by the prior constraints on $A$ and $B$, which are consistent with the background datasets. Given their current level of precision, the existing RSD data do not play a significant role in determining background evolution when combined with DESI-DR2 and SNIa data. Interestingly, when only { DESI-DR2 data are considered, the values of $A$ and $B$ are consistent} with the best-fit $\Lambda$CDM values ($A = 0.315$, $B = 2/3$). Yet, the addition of SNIa data results in noticeable shifts in the mean values of $A$ and $B$, with these shifts being more pronounced with DES-5YR compared to that of PP. This suggests the possibility of presence of new physics or unidentified systematics in the SNIa data.

\item As we mention in the beginning, the background expansion is mainly controlled by two parameters: $A$ and $B$ while the perturbed Universe is controlled by $\gamma$, $\Omega_{m0}$ and $\sigma_{8,0}$. A $\Lambda$CDM Universe, both at background level as well as at perturbed level, demands $B=2/3$, $\gamma=6/11$ and $A=\Omega_{m0}$. All these conditions should be simultaneously satisfied for $\Lambda$CDM Universe. Violations of any of these conditions shows breakdown of $\Lambda$CDM model. Our results, as depicted in Fig \ref{figallcomb}, show that, at present, there is not much correlation between the background parameters and perturbation parameters. This is mainly due to less constraining $f\sigma_{8}$ data. But with more precise data for the growth history of the Universe, as expected from upcoming Stage-IV missions like Euclid \cite{Amendola:2016saw, Euclid:2024yrr}, we expect a larger correlation between background and perturbation parameters and stronger signatures for new physics both at the background and perturbed levels of the Universe. {\color{black} We study such expected signals in section \ref{fc}. Before that, we will discuss specific examples of DDE and MG models that fall within the general framework discussed in this work.}

\end{enumerate}

{\color{black}
\section{Specific models}\label{spmd}
Till now, we have discussed a robust formalism capable of describing both evolving DE and MG scenarios without referring to the exact origin. To further demonstrate the generality and physical relevance of this framework, we now present specific examples corresponding to each of these two scenarios and illustrate how this general formalism naturally accommodates concrete models of evolving DE and MG.

\subsection*{DDE Scenario} 
\begin{itemize}
    \item {\bf Non-interacting DE:} Let us first consider the case of evolving, non-interacting DE. Assuming the matter density $\rho_m=\rho_{m0}(1+z)^3$, the DE density $\rho_{DE}$ at late times can be written as \cite{Mukhopadhyay:2024fch},
\begin{equation}
    \rho_{DE}\sim A(1+z)^\frac{2}{B}+1-A-\Omega_{m0}(1+z)^3.
\end{equation}
For this specific case, the analysis presented in the previous section using our general framework and the resulting constraints on the parameters $A$, $B$, and $\Omega_{m0}$ clearly indicate strong evidence for an evolving DE. In particular, the most compelling evidence comes from the fact that $B = 2/3$ lies outside the $1.8\sigma$ and $2.6\sigma$ confidence levels for the DESI-DR2+PP+RSD and DESI-DR2+DES-5YR+RSD combinations, respectively. 

To better understand the nature of this evolving DE, we examine the corresponding equation of state for DE, which is given by
\begin{equation}
    w_{DE}(z) = \frac{w_T(z) E^2(z)}{E^2(z) - \Omega_{m0}(1+z)^3},
\end{equation}
where \( w_T(z) \) is the total equation of state of the Universe:
\begin{equation}
  w_T(z) = \frac{2A}{3B} \frac{(1+z)^{\frac{2}{B}}}{E^2(z)} - 1.
\end{equation}
We now evaluate \( w_0 \equiv w_{DE}(z=0) \) and \( w_a \equiv \left. \frac{dw_{DE}}{dz} \right|_{z=0} \), and compare them with those from the CPL parametrization of DE \cite{Chevallier:2000qy, Linder:2002et}. To this end, we first constrain the CPL model using the DESI-DR2+PP+RSD and DESI-DR2+DES-5YR+RSD combinations and obtain bounds on the parameters \( w_0,~ w_a,~ \Omega_{m0} \).

We then compare the \( w_0 \)–\( w_a \) parameter space for our specific model—utilizing the constraints already obtained for \( A \), \( B \), and \( \Omega_{m0} \) from these two dataset combinations—with that of the CPL model. This comparison is shown in Fig.~\ref{fig:wcomp_pp_des}. It is evident that, for the current DE model, the allowed parameter space is significantly tighter than that of the CPL model for both dataset combinations and exhibits a different correlation between \( w_0 \) and \( w_a \) compared to the CPL case. Furthermore, most of the allowed region lies within the quintessence regime, in contrast to the CPL model. Similar observations for the current DE model have already been reported in \cite{Mukhopadhyay:2024fch} using the DESI-DR1+PP+RSD combination. For the DESI-DR2+PP+RSD and DESI-DR2+DES-5YR+RSD combinations, we also present the evolution of \( \Omega_m(z) \) and \( \Omega_{DE}(z) \) in Fig.~\ref{fig:Om_pp_des}, and that of \( w_{DE}(z) \) in Fig.~\ref{fig:w_pp_des}.

Therefore, this specific model can also be interpreted as a minimally-coupled scalar field (representing the DE sector) plus pressure-less matter, following the arguments of \cite{Sen:2001xu, Tada:2024znt}, provided it remains confined to the quintessence region. In particular, as shown in \cite{Sen:2001xu}, even when $B = 2/3$, the model does not reduce to $\Lambda$CDM if $A \neq \Omega_{m0}$. In such cases, it still represents an evolving DE component that tracks matter at early times and mimics a cosmological constant at late times.

\begin{figure*}[t]
    \centering
    \includegraphics[width=0.49\linewidth]{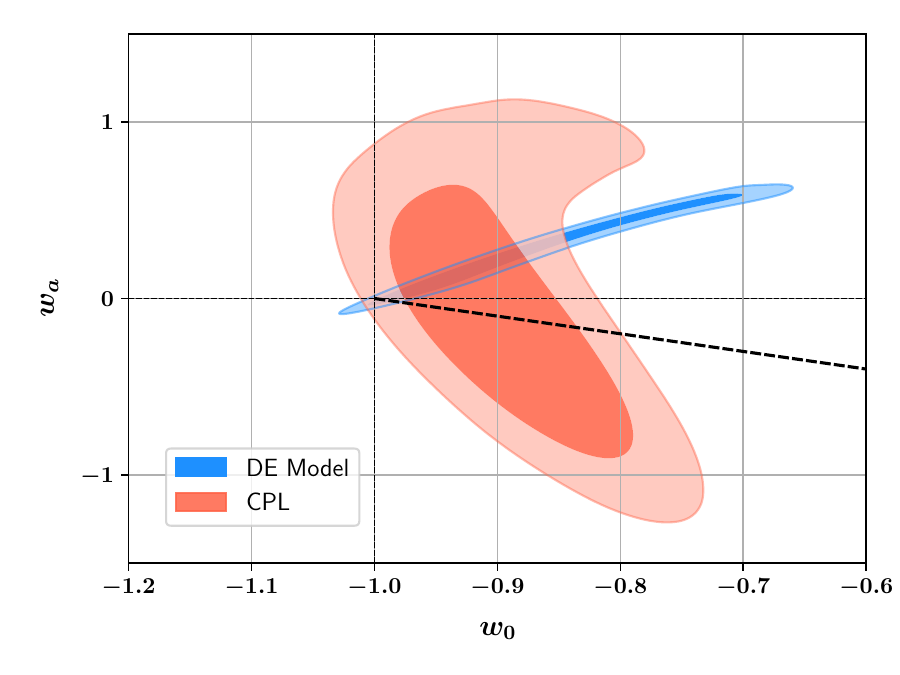}\includegraphics[width=0.49\linewidth]{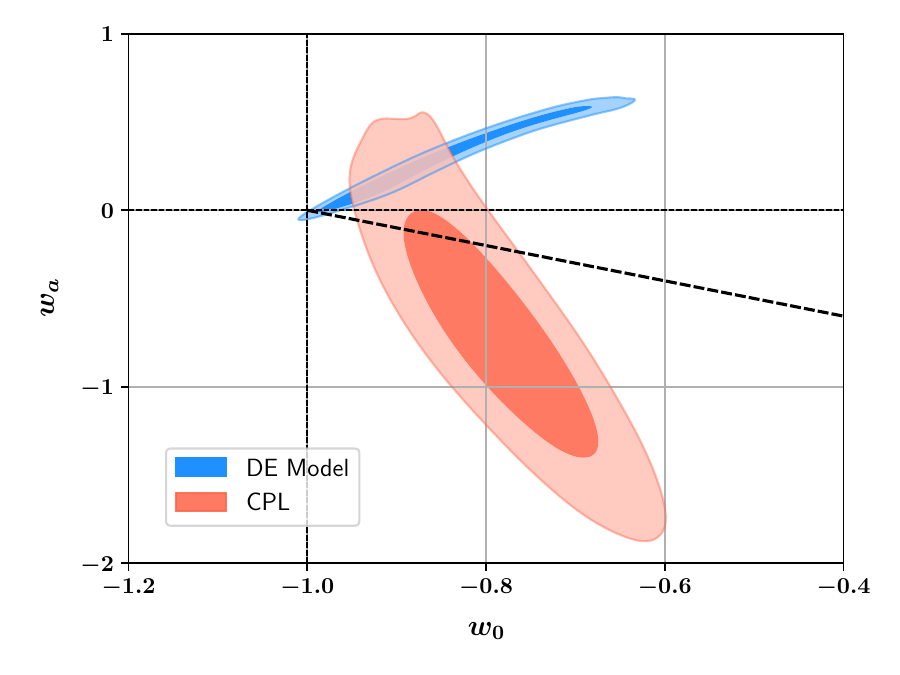}
    \caption{Comparison between the constrained $w_0-w_a$ parameter spaces for the representative non-interacting DE and CPL models obtained from DESI-DR2+PP+RSD (left panel) and DESI-DR2+DES-5YR+RSD (right panel) combinations. Darker and lighter regions represent 68\%, 95\% C.L. respectively. The dashed line represents $w_0+w_a=-1$.}
     \label{fig:wcomp_pp_des}
\end{figure*}

\begin{figure*}[t]
    \centering
    \includegraphics[width=0.49\linewidth]{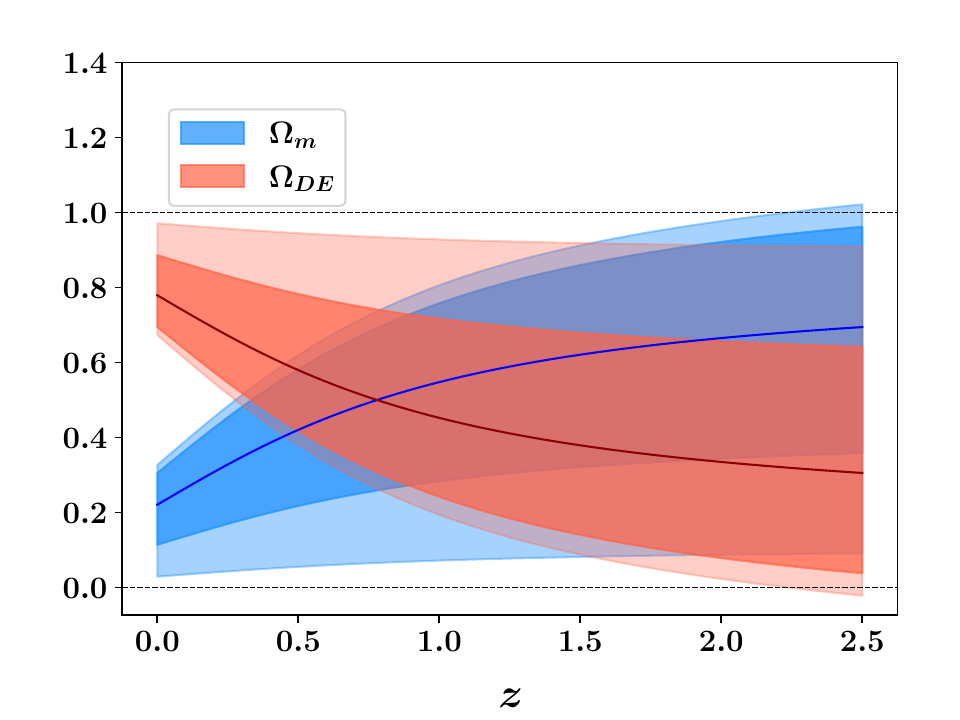}   \includegraphics[width=0.49\linewidth]{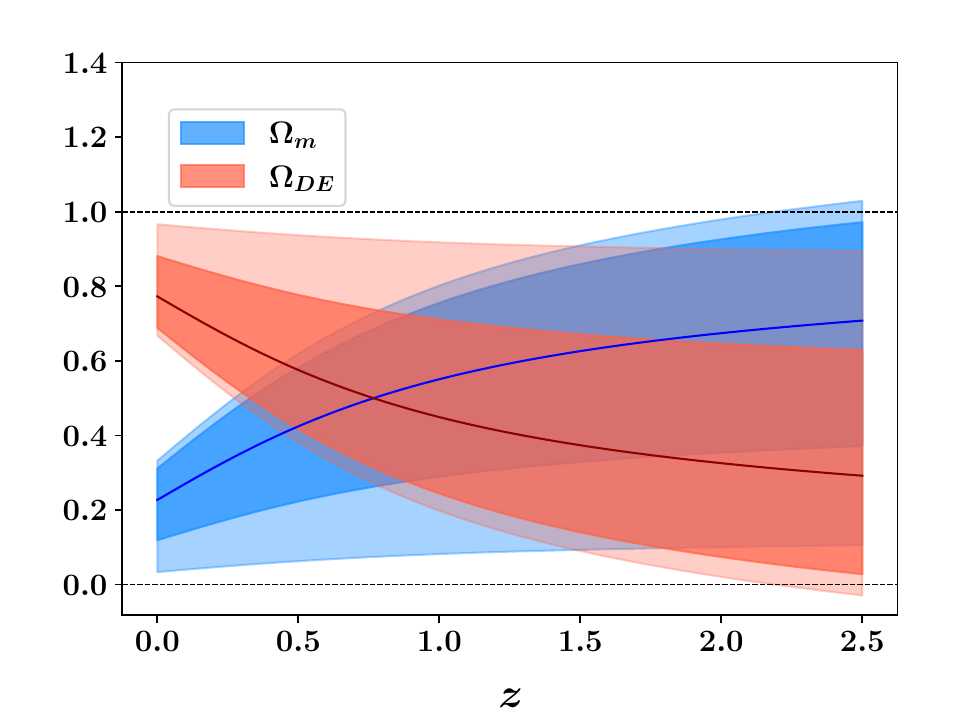}
    \caption{Variations of $\Omega_{DE}$ and $\Omega_{m}$ for the representative non-interacting DE model with redshift z for DESI-DR2+PP+RSD (left panel) and DESI-DR2+DES-5YR+RSD (right panel) combinations. Darker and lighter regions represent 68\%, 95\% C.L. respectively. The solid lines represent the evolutions for best-fit values.}
     \label{fig:Om_pp_des}
\end{figure*}

\begin{figure*}[t]
    \centering
    \includegraphics[width=0.49\linewidth]{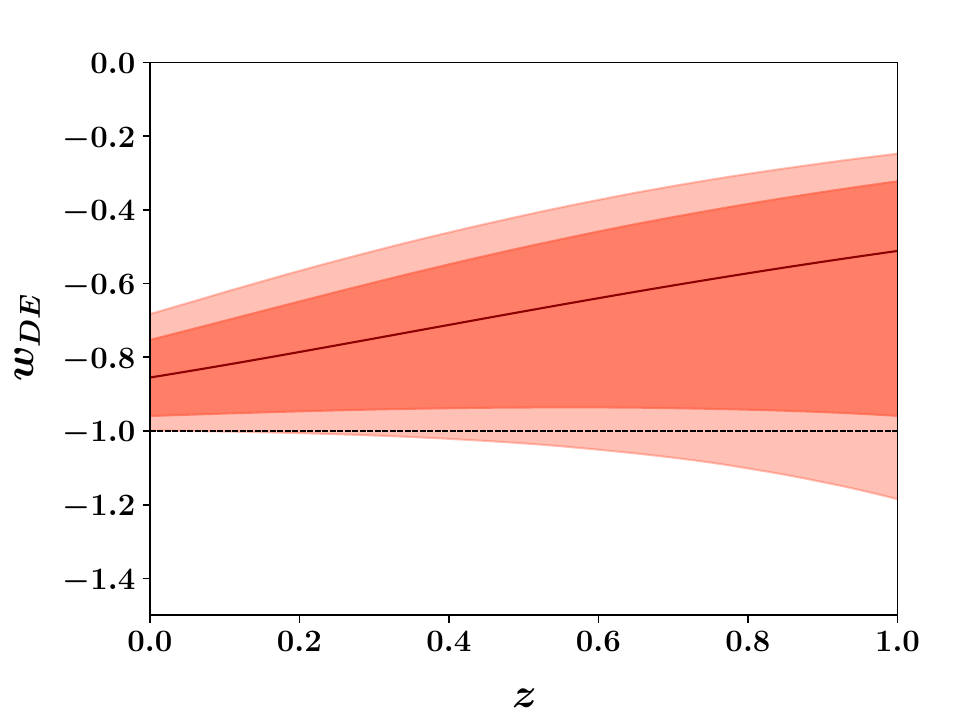} \includegraphics[width=0.49\linewidth]{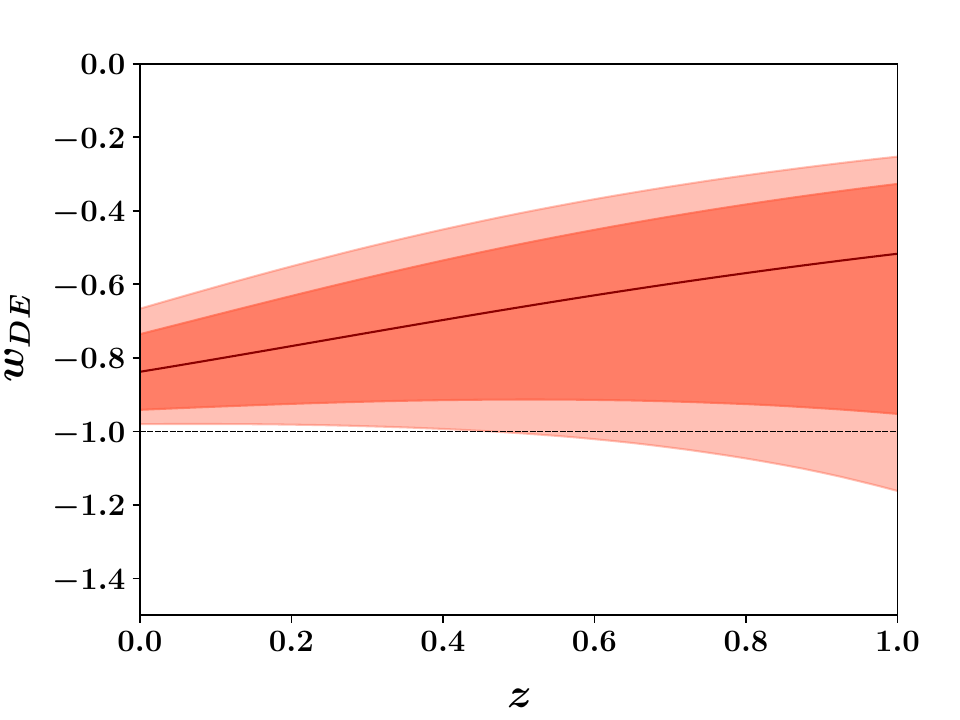}
  \caption{ Variation of $w_{DE}$ for the representative non-interacting DE model with redshift z for DESI-DR2+PP+RSD (left panel) and DESI-DR2+DES-5YR+RSD (right panel) combinations. Darker and lighter regions represent 68\%, 95\% C.L. respectively. The solid lines represent the evolutions for best-fit values.}
     \label{fig:w_pp_des}
\end{figure*}

\item {\bf Interacting DE:}  
Our framework can also accommodate scenarios where pressureless matter interacts with a DDE component (see \cite{Wang:2016lxa} for a recent review on interacting DE models), while still maintaining a DE equation of state \( w_{DE} = -1 \). In this case, the energy densities evolve as follows:
\begin{equation}
\frac{\rho_{m}}{\rho_{c0}} = \frac{2A}{3B}(1+z)^{2/B}, \quad p_{m} = 0,
\end{equation}
\begin{equation}
\begin{split}
\frac{\rho_{DE}}{\rho_{c0}} &= \left(1 - \frac{2}{3B}\right)A(1+z)^{2/B} + (1 - A) \\
&= -\frac{p_{DE}}{\rho_{c0}},
\end{split}
\end{equation}
and the interaction term is given by:
\begin{equation}\label{int}
\frac{Q(z)}{\rho_{c0}} = \left(\frac{2}{B} - 3\right)\frac{2A}{3B}(1+z)^{2/B}H(z),
\end{equation}
which satisfies the following conservation equations:
\begin{equation}
\dot{\rho}_{m} + 3H\rho_m = -\dot{\rho}_{DE} = Q(z).
\end{equation}

Such models have previously been studied in the context of the Generalized Chaplygin Gas (GCG) \cite{Bento:2004uh}. In this sense, our generalized framework can also capture some features of GCG models. It follows from Eq.~\eqref{int} that deviations of \( B \) from \( 2/3 \), as clearly indicated by the analysis in the previous section, may also serve as signatures of an interacting, DDE scenario.

For this model, the redshift evolutions of the fractional matter and DE density parameters are given by:
\begin{equation}
\Omega_{m}(z) = \frac{2A}{3B}\frac{(1+z)^{2/B}}{E^2(z)},
\end{equation}
\begin{equation}
\Omega_{DE}(z) = \left(1 - \frac{2}{3B}\right)\frac{A(1+z)^{2/B}}{E^2(z)} + \frac{1 - A}{E^2(z)}.
\end{equation}
We characterize the interaction using the dimensionless function:
\begin{equation}
\begin{split}
\mathcal{Q}(z) &\equiv \frac{Q(z)}{\rho_{c}(z)H(z)} \\
&= \frac{2A}{3B}\left(\frac{2}{B} - 3\right)\frac{(1+z)^{2/B}}{E^2(z)}.
\end{split}
\end{equation}

Using the constraints on \( A \) and \( B \) obtained from the DESI-DR2+PP+RSD and DESI-DR2+DES-5YR+RSD combinations, we examine the redshift evolutions of \( \Omega_{m}(z) \), \( \Omega_{DE}(z) \), and \( \mathcal{Q}(z) \). The evolutions of \( \Omega_{m}(z) \) and \( \Omega_{DE}(z) \) are shown in Fig.~\ref{fig:Om_pp_des_int}, and that of \( \mathcal{Q}(z) \) in Fig.~\ref{fig:Q_pp_des_int}. Both dataset combinations provide strong evidence for interaction within the interacting DE interpretation of the model, with a more pronounced effect observed for the DESI-DR2+DES-5YR+RSD combination.

\begin{figure*}[t]
    \centering
    \includegraphics[width=0.49\linewidth]{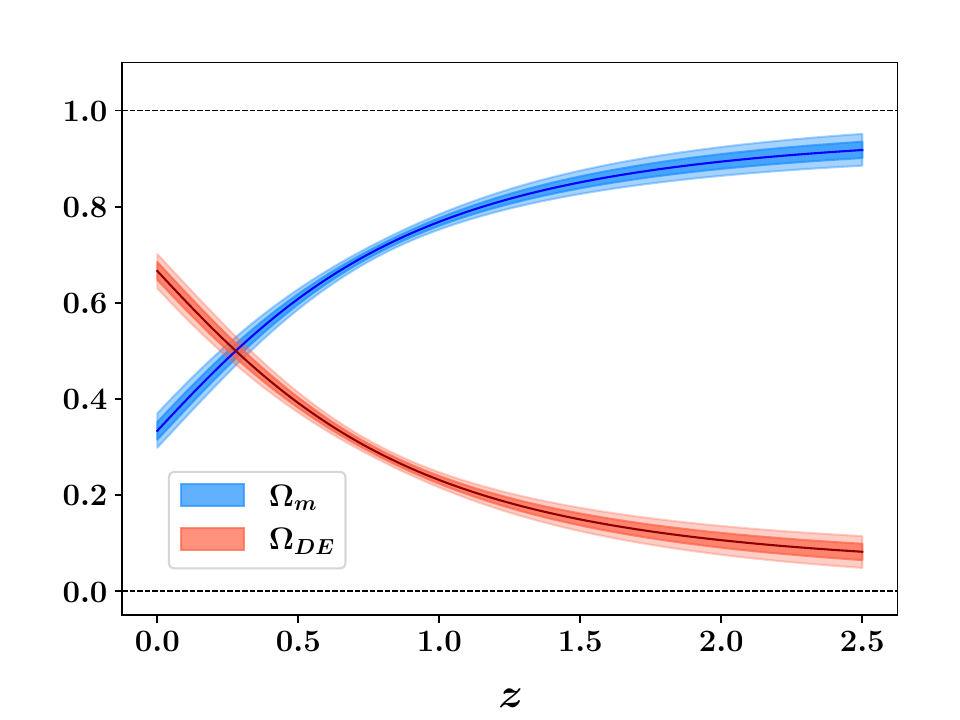}\includegraphics[width=0.49\linewidth]{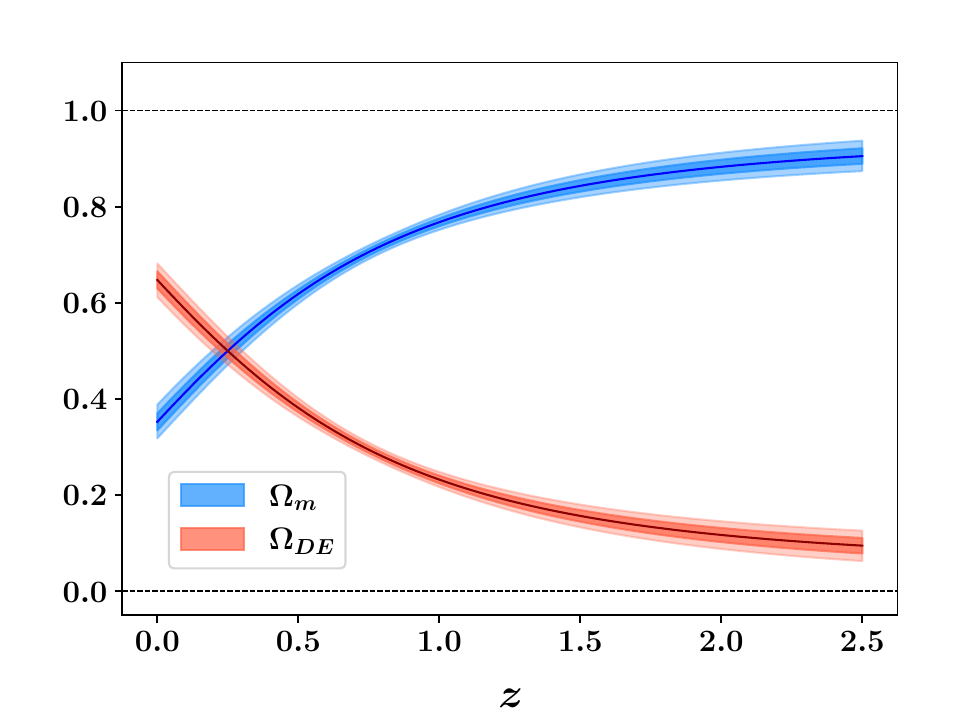}
    \caption{Variations of $\Omega_{DE}$ and $\Omega_{m}$ for the representative interacting DE model with redshift z for DESI-DR2+PP+RSD (left panel) and DESI-DR2+DES-5YR+RSD (right panel) combinations. Darker and lighter regions represent 68\%, 95\% C.L. respectively. The solid lines represent the evolutions for best-fit values.}
     \label{fig:Om_pp_des_int}
\end{figure*}

\begin{figure*}[t]
    \centering
    \includegraphics[width=0.49\linewidth]{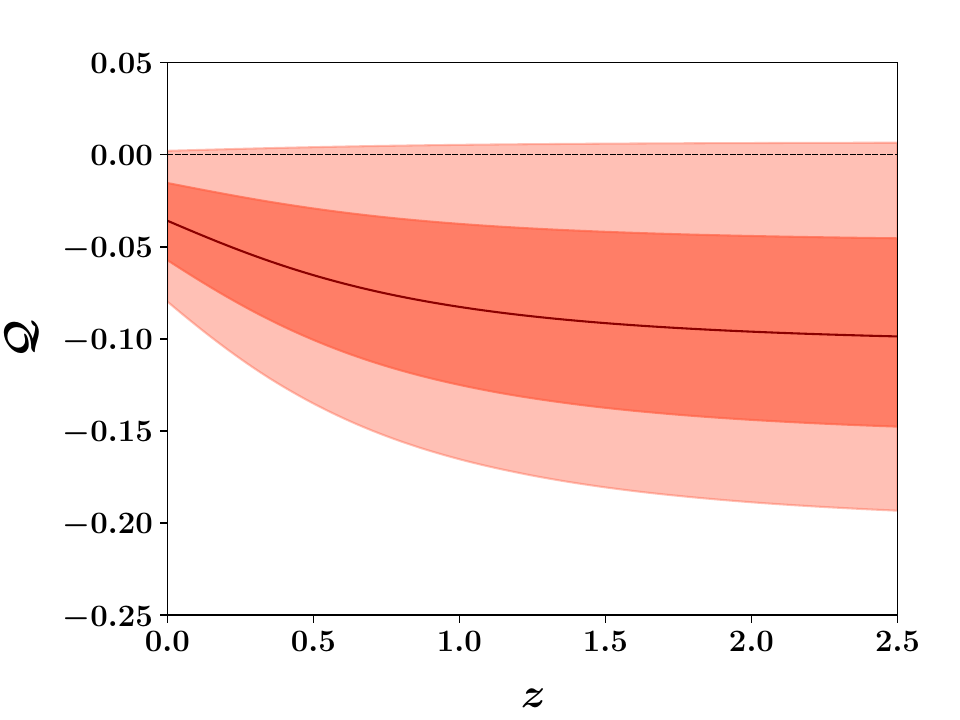} \includegraphics[width=0.49\linewidth]{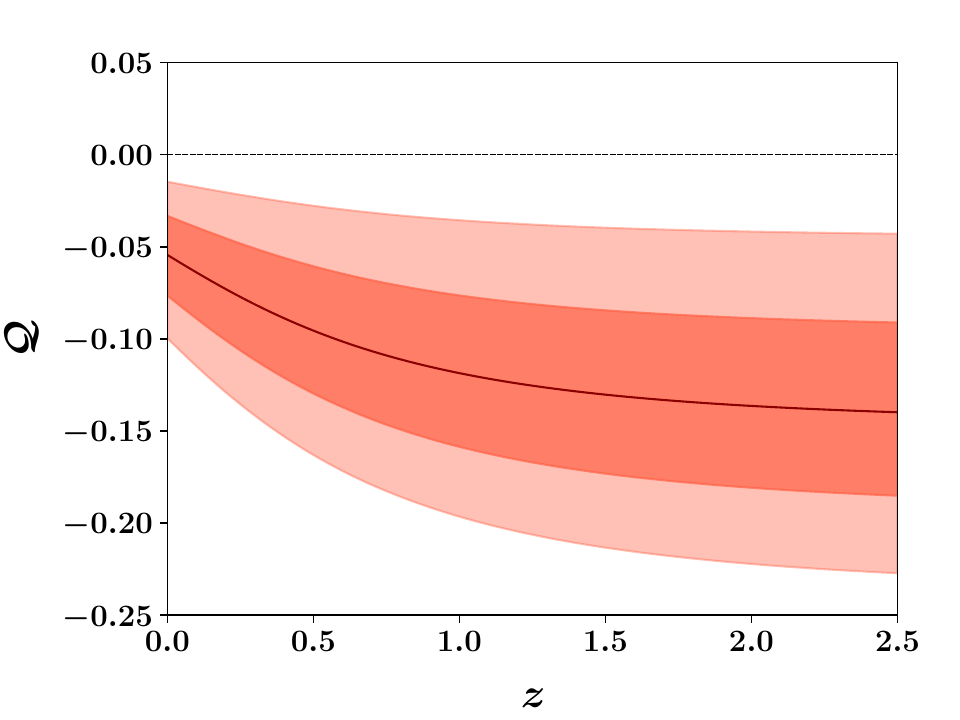}
     \caption{Variation of $\mathcal{Q}$ for the representative interacting DE model with redshift z for DESI-DR2+PP+RSD (left panel) and DESI-DR2+DES-5YR+RSD (right panel) combinations. Darker and lighter regions represent 68\%, 95\% C.L. respectively. The solid lines represent the evolutions for best-fit values.}
     \label{fig:Q_pp_des_int}
\end{figure*}

\end{itemize}

\subsection*{MG Scenario}
Having discussed how evolving DE models fit within our framework, we now turn to MG theories, many of which can also be described by the generalized background evolution \eqref{eqn:H_ABCDM}. As discussed in \cite{Lue:2003ky}, several MG models (including Cardassian models \cite{Freese:2002sq}) can be described by a general modified Friedmann equation, which encapsulates deviations from standard gravity at scales typically much smaller than today’s horizon:
\begin{equation}\label{HMG}
    H^2=H_0^2 ~g(\rho_m),
\end{equation}
where $g(\rho_m)$ is a general function of the matter density $\rho_m$. For our case, using Eq.\eqref{eqn:H_ABCDM}, we can identify,
\begin{equation}
    g(\rho_m)=A\left(\frac{\rho_{m}}{\rho_{m0}}\right)^\frac{2}{3B}+1-A.
\end{equation}
At the background level, we cannot distinguish between such an MG model and the previously discussed evolving DE models, since in both cases the background expansion is governed by the same two parameters, $A$ and $B$. However, the two scenarios lead to different evolution equations for the matter density contrast, which implies that differences should arise when examining the growth of perturbations. A detailed comparison of structure formation in DE-like and MG models, represented by the same Hubble parameter as in Eq.~\eqref{HMG}, is extensively discussed in \cite{Lue:2003ky}. 
We reiterate that the general framework developed in this work naturally accommodates both these scenarios at the level of growth of perturbations through the robust parametrization given in Eq.~\eqref{ffit}, without requiring explicit knowledge of the underlying evolution equations for the matter density contrast. Thus, any potential deviations from GR due to modifications in the gravity sector must be captured by the parameter $\gamma$. 

We have found in the previous section that the current constraints on $\gamma$ are not decisive enough to indicate deviations from $\Lambda$CDM in the perturbation sector, and they remain consistent with the theoretical expectation $\gamma = 6/11$. Therefore, the inferred modification in the background evolution is more likely due to evolving DE rather than MG. In the next section, we discuss how more precise data for the growth history of the Universe from future surveys can play a crucial role in constraining the parameter $\gamma$ significantly and thereby revealing possible signatures of MG. \\

These examples reinforce that our composite framework, governed by the parameters $A$, $B$, and $\gamma$, is flexible enough to describe a wide range of cosmologies, and that observational constraints on these parameters can provide meaningful insights into the viability of both evolving DE and MG models.}

\begin{table*}[t]
\begin{center}
\renewcommand{\arraystretch}{1.35}
\setlength{\tabcolsep}{10pt}
\resizebox{\textwidth}{!}{
\begin{tabular}{l c c c c c c c c}
\hline\hline
\textbf{Datasets} & $\boldsymbol{hr_d}$ & $\boldsymbol{A}$ & $\boldsymbol{B}$ & $\boldsymbol{\Omega_{m0}}$ & $\boldsymbol{\gamma}$ & $\boldsymbol{\sigma_{8,0}}$ & $\boldsymbol{H_0}$ & $\boldsymbol{S_8}$ \\
\hline
DESI-DR2+RSD Forecast & $100.9 \pm 1.3$ & $0.320^{+0.032}_{-0.038}$ & $0.679 \pm 0.019$ & $0.251^{+0.025}_{-0.021}$ & $0.622 \pm 0.041$ & $0.8841^{+0.0097}_{-0.014}$ & $68.61 \pm 0.85$ & $0.807 \pm 0.035$ \\

DESI-DR2+PP+RSD Forecast & $100.13 \pm 0.83$ & $0.343 \pm 0.025$ & $0.690 \pm 0.014$ & $0.262^{+0.021}_{-0.016}$ & $0.643 \pm 0.033$ & $0.8856^{+0.0094}_{-0.013}$ & $68.08 \pm 0.57$ & $0.827^{+0.027}_{-0.024}$ \\

DESI-DR2+DES-5YR+RSD Forecast & $99.32 \pm 0.80$ & $0.369 \pm 0.025$ & $0.702 \pm 0.013$ & $0.274^{+0.019}_{-0.015}$ & $0.666 \pm 0.032$ & $0.8871^{+0.0091}_{-0.012}$ & $67.52 \pm 0.54$ & $0.847^{+0.026}_{-0.022}$ \\
\hline\hline
\end{tabular}
}
\end{center}
\caption{Marginalized $1\sigma$ C.L. limits of the parameters of the composite model across forecasted dataset combinations. $H_0$ (in km Mpc$^{-1}$ s$^{-1}$) is derived from the inverse-distance-ladder method assuming $r_d = 147.09 \pm 0.26$ Mpc \cite{Planck:2018vyg}. $S_8$ is calculated as $S_8 = \sigma_{8,0} \sqrt{\Omega_{m0} / 0.3}$.}
\label{Table_forecast}
\end{table*}

\begin{figure}[t]
    \centering
    \includegraphics[width=\linewidth]{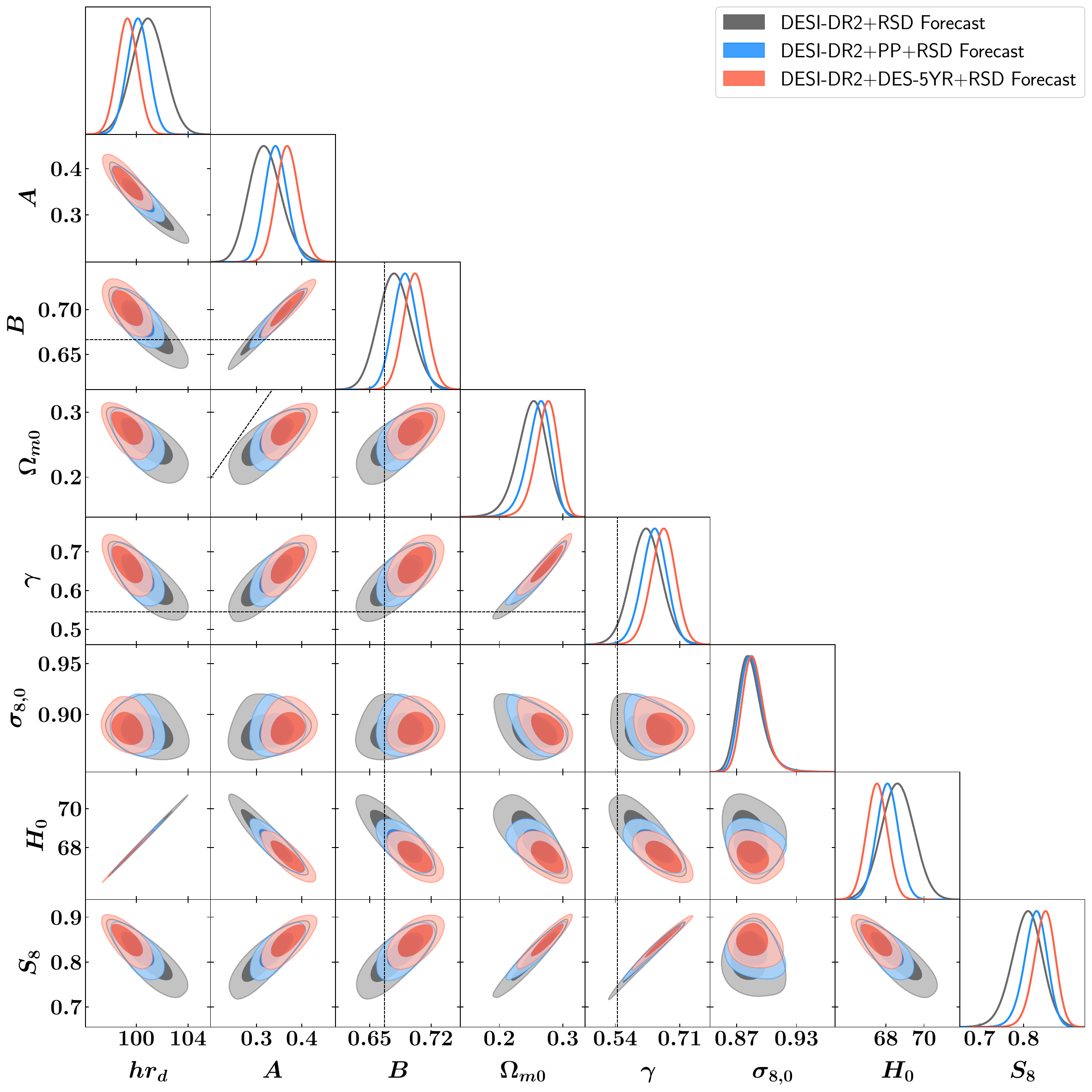}
     \caption{Forecast on one-dimensional marginalized posterior distributions, 2-dimensional contour plots at 68\% and 95\% C.L. limits for parameters of the composite model using combined BAO, SNIa and modified RSD datasets.}
     \label{forecast}
\end{figure}

\section{Forecast}\label{fc}

As discussed in the previous section, the current RSD data do not significantly constrain the background parameters when combined with BAO and SNIa observations. Instead, the constraints on these parameters are entirely determined by the BAO and SNIa datasets. Given this, the role of RSD data is to constrain the parameters governing the growth of perturbations within the background parameter space already constrained by BAO and SNIa. However, the resulting constraints have large confidence intervals and exhibit no correlation with the background parameters.

To examine the potential impact of improved RSD precision, we modify the existing RSD data by reducing the error bars by 10\% at each data point while keeping the central values unchanged. This level of improvement is expected from future Stage-IV surveys such as Euclid \cite{Euclid:2019clj, Euclid:2021xmh}. The resulting constraints and posterior distributions will provide insights into how enhanced RSD precision can refine our understanding of the underlying cosmology.

The results of this analysis are summarized in Fig. \ref{forecast} and Table \ref{Table_forecast}. We observe that the parameters in the perturbation sector are now more tightly constrained, with significantly reduced error bars. Additionally, a clear positive correlation between $A,~B$ and $\gamma,~ \Omega_{m0}$ emerges. In the analysis with the current RSD data, this correlation is largely suppressed (see Fig. \ref{figallcomb}) due to the limited precision of the existing RSD measurements. Thus, RSD data with improved precision is expected to influence not only the constraints on the perturbation sector but also those on the background parameters.

{ Notably, we find that $\gamma = 6/11$ lies outside the $2\sigma$ confidence intervals for both the DESI-DR2+PP+RSD and DESI-DR2+DES-5YR+RSD combinations. Furthermore, the intersection point corresponding to $B = 2/3$ and $\gamma = 6/11$ is excluded at the $2\sigma$ level for these datasets. In addition, the line $A = \Omega_{m0}$ falls outside the $2\sigma$ contours for the DESI-DR2+RSD and DESI-DR2+PP+RSD combinations.} However, it is important to emphasize that in this study, we have kept the central values of $f\sigma_8$ fixed while only reducing the error bars. In contrast, future observational data may yield different central values. Therefore, this forecast should not be interpreted as providing definitive conclusions about the underlying cosmology. Instead, it serves as a guiding framework to assess the potential impact of future RSD data with enhanced precision in constraining both the physics of background evolution and the growth of perturbations.

\section{Conclusions}\label{conc}
{ With the latest DESI-DR2 results together with the largest collection of SnIa data from DES-5YR, we now have a strong indication that the concordance $\Lambda$CDM model does not represent observational data (at least at the background level). With the latest DESI-DR2 data together with DES-5Yr and CMB data from Planck-2018, this is now confirmed at around $4\sigma$ significance level. Given that observables related to background expansion are only sensitive to Hubble parameter $H(z)$ (and hence to total energy density $\rho_{t}$ through Einstein equation), it is still not established whether the non-$\Lambda$CDM behaviour at the background level as observed by various observations, is due to a DDE (minimal or non-minimal we still do not know) or due to modification of gravity. On the other hand, if the new physics is due to modification in the gravity sector itself, that also will have imprint in the growth of structures.  Hence it is essential to disentangle and independently constrain the growth of perturbations from the background dynamics. 

In this work, we have incorporated the new physics in the background evolution through the modified scale factor and in the perturbed sector through the growth rate without assuming any specific dark energy or modified gravity model.}
 
 At the background level, the deviations from $\Lambda$CDM model are described by two parameters, $A$ and $B$, which reduce to $\Omega_{m0}$ and $2/3$, respectively, in the standard $\Lambda$CDM scenario. To probe deviations in the growth of structures, we introduce a fitting function for $f\sigma_8$, which parameterizes possible departures from $\Lambda$CDM through the growth index $\gamma$. In the standard $\Lambda$CDM model, $\gamma=6/11$. With this, we constrain the background and perturbation parameters using SN-Ia and BAO data, as well as growth rate measurements from RSD.

We find that the background parameters, \( A \) and \( B \), are primarily constrained by SN-Ia and BAO observations, while RSD data predominantly constrain the parameters associated with the perturbation sector, namely \( \Omega_{m0} \), \( \gamma \), and \( \sigma_{8,0} \), with minimal impact on \( A \) and \( B \). { Our analysis provides significant evidence for deviations from the \( \Lambda \)CDM model in the background evolution. Specifically, using the DESI-DR2+PP+RSD and DESI-DR2+DES-5YR+RSD combinations, we find that \( B=2/3 \) is excluded at \( 1.8\sigma \) and \( 2.6\sigma \) CL, respectively. 
On the other hand,  the best-fit value of \( \gamma \) in our analysis is close to the \( \Lambda \)CDM expectation indicating no deviation from $\Lambda$CDM in the perturbed Universe, with the current growth data. Additionally, we get lower allowed value for $\Omega_{m0}$ from growth data, compared to the Planck-2018 constraint on $\Omega_{m0}$ as shown in  Fig. \ref{figallcomb} as well as in Table \ref{Table_gm}.}

Therefore, with the current data for both background evolution as well as for growth of structures, our results indicate that while the growth of structures is consistent with the predictions of GR, the data favour a lower value of \( \Omega_{m0} \) compared to \( \Lambda \)CDM. This, in turn, suggests that observational data exhibit a preference for modifications in the background evolution, likely due to an evolving DE component, rather than modifications to gravity itself.

Motivated by our findings with the current RSD data that it does not significantly constrain the background parameters when combined with SN-Ia and BAO datasets, and no apparent correlation is observed between the parameters governing the background and perturbations, we investigate the potential impact of more precise RSD data from future surveys on constraining the underlying cosmology. Our analysis suggests that even a 10\% improvement in precision could play a crucial role in detecting possible new physics in both the perturbation and background sectors. This is because a correlation between the background and perturbation parameters is expected to emerge, which is currently suppressed due to the limited precision of existing RSD data.

{ In conclusion, we present a unified model independent approach to look for new physics both in the background evolution as well as in the growth of structures. {\color{black} We have also demonstrated that our unified framework naturally accommodates both DDE and MG scenarios by discussing specific representative models. While the background dynamics can be universally described by the parameters \( A \) and \( B \), the growth of perturbations—captured through the parameter \( \gamma \)—acts as a discriminator between these two classes of models.}
 With the upto-date data from DESI-DR2 for the redshift space distortion are forthcoming as well as we expect the new measurements of both BAO and redshift space distortion from Euclid, this approach can be very useful to probe new physics beyond $\Lambda$CDM model.}

\section*{Acknowledgments}

SGC acknowledges funding from the Anusandhan National Research Foundation (ANRF), Govt. of India, under the National Post-Doctoral Fellowship (File no. PDF/2023/002066). PM acknowledges funding from ANRF, Govt. of India, under the National Post-Doctoral Fellowship (File no. PDF/2023/001986). AAS acknowledges the funding from ANRF, Govt. of India, under the research grant no. CRG/2023/003984. We acknowledge the use of the HPC facility, Pegasus, at IUCAA, Pune, India.

\bibliography{ref.bib}

\end{document}